\documentclass[superscriptaddress,runinaddress,10pt, twocolumn,showpacs,showkeys,floatfix]{revtex4}

\usepackage{doi}
\usepackage{hyperref}
\hypersetup{
  colorlinks=true,        
  linkcolor=blue,         
  citecolor=cyan,         
}

\usepackage{dcolumn}
\usepackage{bm}
\usepackage[mathlines]{lineno}

\usepackage{palatino}
\usepackage{mathrsfs}
\usepackage{hyperref}
\usepackage{xcolor}
\usepackage{float}
\usepackage{ulem}
\usepackage{lipsum}
\usepackage{graphicx}

\begin{document}

\title{Testing Cosmic Censorship Conjecture for Extremal and Near-extremal $(2+1)-$dimensional MTZ Black Holes}
\author{Koray D\"{u}zta\c{s}}
\email{ koray.duztas@ozyegin.edu.tr}
\affiliation{Department of Natural and Mathematical Sciences,
\"{O}zye\u{g}in University, 34794 \.{I}stanbul, Turkey}

\author{Mubasher Jamil}\email{mjamil@zjut.edu.cn }
\affiliation{Institute for Theoretical Physics and Cosmology, Zheijiang University of Technology, Hangzhou 310023, China}
\affiliation{United Center for Gravitational Wave Research, Zheijiang University of Technology, Hangzhou 310023, China}
\affiliation{Department of Mathematics, School of Natural Sciences (SNS), National
University of Sciences and Technology (NUST), H-12, Islamabad 44000, Pakistan}
\affiliation{Canadian Quantum Research Center 204-3002 32 Ave Vernon, BCV1T 2L7, Canada}

\author{Sanjar Shaymatov} \email{sanjar@astrin.uz}
\affiliation{Ulugh Beg Astronomical Institute, Astronomicheskaya 33, Tashkent 100052, Uzbekistan}
\affiliation{National University of Uzbekistan, Tashkent 100174, Uzbekistan}
\affiliation{Tashkent Institute of Irrigation and Agricultural Mechanization Engineers,\\ Kori Niyoziy 39, Tashkent 100000, Uzbekistan }

\author{Bobomurat Ahmedov} \email{ahmedov@astrin.uz}
\affiliation{Ulugh Beg Astronomical Institute, Astronomicheskaya 33, Tashkent 100052, Uzbekistan} 
\affiliation{National University of Uzbekistan, Tashkent 100174, Uzbekistan}
\affiliation{Tashkent Institute of Irrigation and Agricultural Mechanization Engineers,\\ Kori Niyoziy 39, Tashkent 100000, Uzbekistan }

\begin{abstract}
We test the validity of the weak cosmic censorship conjecture for the ($2+1$)-dimensional charged anti-de Sitter  black hole solution, which was derived by Martinez, Teitelboim, and Zanelli (MTZ). We first construct a thought experiment by throwing  test charged particles on an extremal MTZ black hole. We derive that extremal ($2+1$) dimensional black holes can be overcharged by test particles, unlike their analogues in ($3+1$) and higher dimensions. Nearly-extremal MTZ black holes can also be overcharged, by a judicious choice of energy and charge for the test particles when  the second order effects are ignored. We also incorporate the second order effects for nearly extremal MTZ black holes by adapting the method developed by Sorce and Wald. However it turns out that the second order effects cannot compensate for the overcharging of MTZ black holes. 
\end{abstract}

%
\vspace{2pc}
%

\maketitle
\section{Introduction}

Penrose and Hawking have shown that gravitational collapse of massive objects leads inevitably to curvature singularities \cite{Hawking-Penrose70,Penrose69}. The {\it weak cosmic censorship conjecture} (WCCC) proposed by Roger Penrose asserts that naked spacetime singularities (not hidden behind an event horizon) must be forbidden in a physical universe (see \cite{Jacobson10a} for a review). The conjecture exists so far without concrete proof and is generally considered as a fundamental law of general relativity. The validity of the CCC is consistent with the second law of thermodynamics and the energy conditions. However turning a black hole into a naked singularity ultimately destroys some of the energy conditions and the Hawking-Bekenstein's famous entropy-area law. If the conjecture is violated somehow, the exposed singularity may provide a window to test the theories of quantum gravity. The validity of the conjecture has been tested via numerous Gedanken experiments for extremal and near-extremal black holes in the literature. The first Gedanken experiment in this vein was constructed by Wald. He showed that particles which carry sufficient charge or angular momentum to overcharge or overspin an extremal  Kerr-Newman black hole are not absorbed by the black hole~\cite{Wald74b}. This result was also generalised to scalar test fields~\cite{Semiz11,Toth12}. 

Later, Hubeny proposed an alternate approach where one starts with a nearly extremal black hole instead of an extremal one~\cite{Hubeny99}. She showed that it could be possible to overcharge a nearly extremal electrically charged Reissner-Nordstr\"{o}m black hole by using tailored charged particles. This approach was also applied to Kerr and Kerr-Newman black holes~\cite{Jacobson09,Saa11}. Later backreaction effects were considered for these cases to prevent the horizon from being destroyed~\cite{Barausse10,Isoyama11,Zimmerman13}. It is observed that backreaction effects usually prevent the formation of naked singularities. The same is also true for magnetic field however, de Felice and Yunqiang showed that an extremal Reissner-Nordstrom black hole may be turned into a Kerr-Newman naked singularity after capturing an electrically neutral spinning body \cite{deFelice01}. Recently, similar conclusions have been drawn by over-charging the higher dimensional nearly extremal charged black holes using the new version of gedanken experiment \cite{Ge18}. The same question was analysed for test fields instead of particles. Similar results were found for Kerr black holes interacting with bosonic test fields~\cite{Duztas13,Duztas14,Natario16}. However, the interaction with massless Dirac fields can lead to the destruction of extremal black holes~\cite{Duztas15,Toth16}. The effect of Hawking radiation was also incorporated in the problems involving test fields~\cite{Duztas16a}. There is also an investigation that suggests that a test magnetic field would serve as CCC, preventing black hole horizon from being destroyed~\cite{Shaymatov15} as well as the same is true even for back-reaction effect of the magnetic field~\cite{Shaymatov19b}.

The quantum connection with CCC was analysed in \cite{Matsas07PRL,Hod08PRL,Richartz08,Hod08PLB,Matsas09,Richartz11,Semiz15}. The validity of WCCC was also investigated for the asymptotically anti-de Sitter case~\cite{Zhang14,Rocha14,Gwak16JCAP,Gwak16PLB}. Further the CCC has been considered also in various framework, for example the Gauss-Bonnet AdS black hole~\cite{Zeng19a}, Born-Infeld AdS black hole~\cite{Zeng19b}, the quintessence AdS black hole~\cite{He19} and string analog of Reissner-Nordstr\"{o}m black holes \cite{Duztas19}. No violation of the weak cosmic censorship conjecture was found around the five-dimensional Myers-Perry black holes for the non-linear particle accretion~\cite{An18}. However, Shaymatov et. al. showed that all higher dimensional ($>4$) rotating black holes having only single rotation would always obey the weak cosmic censorship conjecture even in the liner order regime~\cite{Shaymatov19a,Shaymatov20a}. The same result was obtained in the case of five dimensional charged rotating black hole with single rotation -- the CCC is strongly respected when angular momentum dominates over charge~\cite{Shaymatov19c}.    
Siahaan showed that if one ignores the self-force, self-energy and radiative effects, an extremal or a near-extremal Kerr-Sen black hole can turn into a naked singularity when it captures charged and spinning massive particles \cite{Siahaan16}. It was also shown that test fields can destroy the event horizons of extremal and nearly extremal Kerr-Taub-NUT black holes~\cite{Duztas18}. Following~\cite{Gao-Wald01}, Sorce and Wald have recently published new versions of their original thought experiment~\cite{Wald18,Sorce-Wald17}.

In literature, the first test of WCCC  for the case of $(2+1)-$ dimensional extremal spinning Banados, Teitelboim, Zanelli (BTZ) black holes was performed by Rocha and Cardoso~\cite{Rocha11}, where they concluded that BTZ black holes cannot be overspun. Later it was shown that overspinning is possible if one starts with a nearly extremal BTZ black hole instead \cite{Duztas16}. The charged black hole solution for the $(2+1)-$ dimensional case was derived by Martinez, Teitelboim and Zanelli \cite{Martinez00}. In this work we are motivated by Hubeny to test the validity of WCCC in the case of massive charged particles interacting with MTZ black holes carrying electric charge but no spin. In the work of Hubeny and its recent generalization to higher dimensional black holes by Revelar and Vega \cite{Revelar-Vega17}, the authors concluded that nearly extremal black holes can be overcharged, though extremal black holes cannot. Here we answer the question whether this can also be generalised to the $(2+1)-$ dimensional case.
We shall use the conventions $c=G=1$, and ignore back-reaction effects. All indices are taken to run from 0 to 2.

\section{Overcharging ($2+1$) dimensional black holes}
We start with the Einstein-Hilbert-Maxwell action:
\begin{equation}
\mathcal S=\int d^3x \sqrt{-g}\Big(  \frac{R-2\Lambda}{16\pi} -\frac{1}{4}F_{\mu\nu}F^{\mu\nu} \Big)\, .
\end{equation}
Here $F_{\mu\nu}$ is tensor of electromagnetic field, while $R$ is the scalar curvature of the spacetime. In the following, we shall fix $\Lambda=-l^{-2}$ being equal to unity. After solving the field equations in Hamiltonian form with the assumptions of rotational symmetry and time independence, Martinez-Teitelboim-Zanelli (MTZ) obtained the following solution representing a charged black hole without angular momentum~\cite{Martinez00}
\begin{equation}
ds^2=-f(r)dt^2 + \frac{dr^2}{f(r)}+r^2d\phi^2,
\end{equation}
where the metric function 
\begin{equation} \label{f(r)}
f(r)=r^2-M-\left( \frac{Q}{2}\right)^2\ln(r^2)\, ,
\end{equation}
with $M$ being MTZ black hole mass and $Q$ being the total electric charge of black hole. The function $f(r)$ has a minimum at $r_{\rm{min}}=Q/2$. The value of this function at its minimum is
\begin{equation}
f(r_{\rm{min}})=-M+\left( \frac{Q}{2} \right)^2\left[ 1- \ln\Big( \frac{Q}{2}\Big)^2 \right].
\end{equation}
There are three possibilities to characterize the spacetime: If $f(r_{\rm{min}})=f(Q/2)<0$, there exist two roots of $f(r)$. Then we have a usual black hole with $r_+$, and $r_-,$ as the inner and outer horizons. If $f(r_{\rm{min}})=f(Q/2)=0$, the two roots coincide and we have an extremal black hole. If $f(r_{\rm{min}})=f(Q/2)>0$, there are no real roots of $f(r)$, hence we have a naked singularity. The case of extremal black holes corresponds to $f(Q/2)=0$. Since $f(r_+)=0$ by definition, for an extremal black hole, we have $r_+=Q/2$.

In Wald type Gedanken experiments we start with an extremal or a nearly extremal black hole satisfying the relevant equations. Then, we send in test particles or fields from infinity. After test particles or fields interact with the black hole the space-time settles to its final configuration, with new parameters of mass, angular momentum, charge etc. Finally we check if the final configuration of parameters represent a black hole or a naked singularity. Here, we test the validity of WCCC for a MTZ black hole  interacting with test charged particles. The general equations of motion of a test particle of mass $m$ and charge $q$ in a curved background are given by
\begin{equation}
\ddot{x}^\mu+ \Gamma^{\mu}_{\rho \sigma}\dot{x}^{\rho} \dot{x}^{\sigma}=\frac{q}{m}F^{\mu \nu}\dot{x}_{\nu},
\end{equation}
which can be derived from the Lagrangian
\begin{equation}
\mathcal{L}=\frac{1}{2}mg_{\mu \nu}\dot{x}^{\mu}\dot{x}^{\nu}+ qA_{\mu}\dot{x}^{\mu},
\end{equation}
where ${\bf A}=-Q\ln(r) dt$, i.e. $A_0=-Q\ln(r)$, and the dot denotes derivative with respect to the affine parameter.
The non-vanishing component of the so-called Faraday tensor (also known as the Maxwell tensor) after evaluation yields:  
\begin{eqnarray}\label{Eq:Far}
F_{rt}&=& - \frac{Q}{r}\, . 
\end{eqnarray}
To find the coordinate components of electromagnetic field one needs to consider the proper observer for which the
three-velocity components are defined by 
\begin{eqnarray}
\label{Eq:zamo_con} & \dot{x}^{\mu}
=\left\{\left({r^2-M-\left( \frac{Q}{2}\right)^2\ln(r^2)}\right)^{-1/2},0,0\right\}\
, \\
\label{Eq:zamo_cov} &\dot{x}_{\mu}=
\left\{-\left({r^2-M-\left( \frac{Q}{2}\right)^2\ln(r^2)}\right)^{1/2},0,0\right\}\, .
\end{eqnarray}
It is then straightforward to evaluate the non-vanishing orthonormal component of the electromagnetic field as 
\begin{eqnarray}
\label{Er}  E_{\hat r}
&=&e^{r}_{\hat r}E_{r}\equiv-\frac{Q}{r} \, , 
\end{eqnarray}
where $\bf{e}_{\hat r}$ is the orthonormal tetrad. In the limit $r\rightarrow \infty$ the radial component of electric field $ E_{\hat r}\rightarrow 0$. This clearly indicates that the electric field is purely radial around the MTZ black hole, so the charged particle gets interacted with the fields in the black hole vicinity.

{Let us then define the associated conserved quantities such as the energy and angular momentum of the particle around the black hole as follows:}
\begin{equation}
E=-\frac{\partial \mathcal{L}}{\partial \dot{t}}=mf(r)\dot{t}+qQ\ln(r), \label{E}
\end{equation}
and
\begin{equation}
L=\frac{\partial \mathcal{L}}{\partial \dot{\phi}}=mr^2\dot{\phi} .\label{L}
\end{equation}
Usually one needs to evaluate the energy equation (\ref{E}) with (\ref{L}) and the condition $-1=g_{\mu \nu}\dot{x}^{\mu}\dot{x}^{\nu}$ to find an expression for the minimum energy so that the particle crosses the horizon and the radial equation of motion as well 
\begin{eqnarray}\label{radial}
\dot{r}^2=\frac{1}{m^2}\left[E-qQ\ln(r)\right]^2-f(r)-\frac{f(r)}{m^2r^2}L^2\, ,
\end{eqnarray}
where the third term on the right hand side vanishes for a freely falling particle.
In this case there are no off-diagonal terms in the metric. The equation (\ref{E}) is sufficient for us to conclude that, at $r=r_+$, the energy of the particle at the horizon is $E_{\rm{min}}=qQ\ln(r_+)$. On the other hand, using $f(r_+)=0$ for Eq.~(\ref{radial}), we get the lower bound of energy as
\begin{equation}
E\geq E_{\rm{min}}= qQ\ln(r_+).
\label{eminopt}
\end{equation}
The above constraint is consistent with the fact that $\dot r ^2>0$ for all $r\geq r_+$ \cite{Hubeny99}. If the energy of the particle is lower than $E_{\rm{min}}$ it cannot cross the horizon, i.e. it will not be absorbed by the black hole.

\subsection{Overcharging extremal MTZ black hole}

For a black hole solution, we require $f(r_{\rm{min}})\leq0$, i.e.
\begin{equation}\label{Eq:extremal}
\delta\equiv M-\left( \frac{Q}{2} \right)^2\left[ 1- \ln\Big( \frac{Q}{2}\Big)^2 \right] \geq 0\, .
\end{equation}
Note that the function $( Q/2)^2 [ 1- \ln (Q/2)^2 ]$ vanishes at $Q=0$ and $(Q/2)^2=e$, but has a maximum at $(Q/2)^2=1$ (or $Q=2$), which is equal to 1.  Thus, if $M>1$, $\delta$ is always larger than zero so we have an ordinary black hole with $r_+$ and $r_-$. 

We have to consider the cases $M<1$ to overcharge extremal black holes.  Among these black holes we should also exclude the solutions with $(Q/2)^2>1$, since the function $(Q/2)^2[1-\ln (Q/2)^2]$ decreases after its maximum point $(Q/2)^2=1$. After the maximum point $(Q/2)^2=1$, when the charge of the black hole increases, $\delta$ as defined in (\ref{Eq:extremal}) also increases; i.e. the MTZ black hole is driven away from extremality. Therefore the solutions with $M<1$ and/or $(Q/2)^2>1$ are not of interest in an attempt to overcharge a  MTZ black hole. 

Let us start with an extremal black hole with $\delta_{\rm{in}}=0$. We perturb this black hole with test particles which have energy $E$ and charge $q$. Notice that, since $M<1$, and $(Q/2)^2<1$, we have $r_+<1$, thus the minimum energy to have the particle absorbed by the black hole is {\it negative}. We derive the maximum energy for the particle by demanding that the final configuration of the space-time parameters represents a naked singularity, i.e.
\begin{eqnarray}
 \delta_{\rm{fin}}&=&(M+\delta E)-\left( \frac{Q+\delta Q}{2}\right)^2\nonumber\\&& + \left( \frac{Q+\delta Q}{2}\right)^2\ln\left[ \left( \frac{Q+\delta Q}{2}\right)^2 \right]<0. \label{deltafin}
\end{eqnarray}
Under a reasonable assumption that particle's charge is considerably smaller than black hole's charge, let us choose $\delta Q=\epsilon Q$ (where $\epsilon \ll 1$) for the charge of test particles so that the test particle approximation is not violated. In that case
\[
\ln\left[ \left( \frac{Q+\delta Q}{2}\right)^2 \right]=\ln\left[ \left( \frac{Q}{2}\right)^2\right]+\ln \left[ ( 1+\epsilon)^2 \right].
\]
For small $\epsilon$, we can make the expansions
$\ln(1+\epsilon)\simeq\epsilon - \frac{\epsilon^2}{2}+O(\epsilon^3),$ and $\ln(1+\epsilon)^2\simeq2\epsilon -\epsilon^2.$
Than, retaining terms up to second order
\[
\ln\left[ \left( \frac{Q+\delta Q}{2}\right)^2 \right]=\ln\left[ \left( \frac{Q}{2}\right)^2 \right] +2\epsilon -\epsilon^2.
\]
We can rewrite Eq. (\ref{deltafin}) as
\begin{eqnarray}
&& M+\delta E-\left( \frac{Q}{2}\right)^2-\epsilon^2 \left( \frac{Q}{2}\right)^2 -2\epsilon \left( \frac{Q}{2}\right)^2 + \nonumber \\
&& \left\{ \left( \frac{Q}{2}\right)^2+\epsilon^2 \left( \frac{Q}{2}\right)^2+2\epsilon\left( \frac{Q}{2}\right)^2\right\}
\nonumber\\&&\times \left\{\ln\left[ \left( \frac{Q}{2}\right)^2 \right] +2\epsilon -\epsilon^2 \right\}<0 .
\label{deltafin1}
\end{eqnarray}
Working up to second order in $\epsilon$ and using $\delta_{\rm{in}}=0$, we arrive at
\begin{equation}
\delta E<\delta E_{\rm{max}}=-(\epsilon^2+2\epsilon)\left( \frac{Q}{2}\right)^2\ln\left[\left( \frac{Q}{2}\right)^2\right]-2\epsilon^2 \left( \frac{Q}{2}\right)^2. \label{deltafin2}
\end{equation}
The right hand side of the inequality is positive for $(Q/2)^2<e^{(-2\epsilon)/(\epsilon +2)}$. We choose positive energy for  particles at infinity. Thus, an extremal black hole can be overcharged with two choices. We choose the charge of our incoming particles $\delta Q=\epsilon Q$, and their energies are in the range $0<\delta E<  \delta E_{\rm{max}}$, where $ \delta E_{\rm{max}}$ is given by (\ref{deltafin2}).

\begin{figure}
\centering
\includegraphics[width=0.45\textwidth]{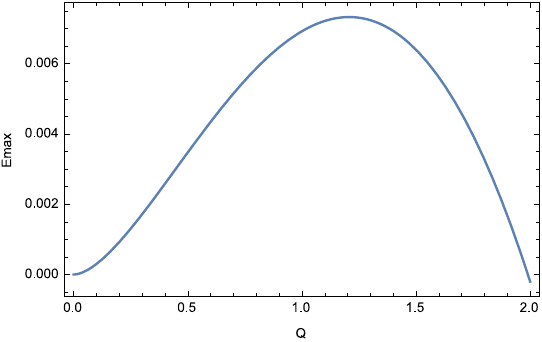}\caption{\label{fig1} Graph of maximum energy $\delta E_{\rm{max}}$ against black hole charge $Q$. Here we choose $\epsilon=0.01$.}
\end{figure}

\begin{figure}
\centering
\includegraphics[width=0.45\textwidth]{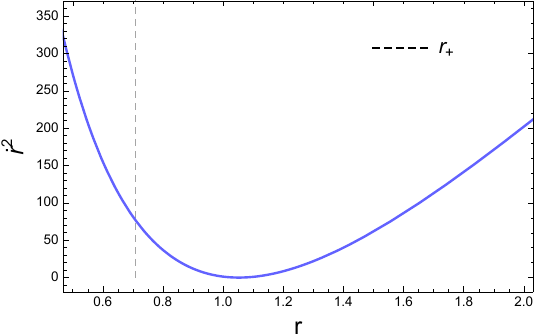}
\caption{\label{fig2} Graph of radial dependence of the motion of
charged particle falling in an extremal MTZ black hole. Following the numerical example we choose parameters.}
\end{figure}

Fig.~\ref{fig1} shows the plot of the dependence of maximum energy from black hole electric charge. Since $\delta E_{\rm{min}}$ is negative, for each specific $Q$ we can choose any value of $\delta E$ under the curve, and $\delta Q=\epsilon Q$ to overcharge the black hole. For a numerical example let us start with an extremal black hole with $(Q/2)^2=0.5$. Since $\delta_{\rm{in}}=0$, $M=0.8466$ up to four significant digits. Letting $\delta Q=\epsilon Q$ with $\epsilon=0.01$, we get $E_{\rm{max}}=0.006866$. We see that $\delta E_{\rm{max}}\lesssim M\epsilon $, so that the test particle approximation is not violated. Let us choose $\delta E=0.001<E_{\rm{max}}$ for falling in test particle.  Then $\delta_{\rm{fin}}$ is given by
\begin{eqnarray}
 \delta_{\rm{fin}}&=&M+\delta E-\left( \frac{Q+\delta Q}{2}\right)^2  \nonumber\\&&+\left( \frac{Q+\delta Q}{2}\right)^2\ln\left[ \left( \frac{Q+\delta Q}{2}\right)^2 \right]\nonumber\\&=&-0.00584.
\label{exext}
\end{eqnarray}
The negative sign indicates that the extremal black hole is overcharged into a naked singularity. The radial motion of a freely falling charged test particle is also shown in Fig.~\ref{fig2}. As can be seen from Fig.~\ref{fig2}, $\dot r ^2>0$ is always satisfied around black hole horizon $r_{+}$, and hence the charged particle with appropriate parameters crosses the horizon as there appears no turning point.

\subsection{Overcharging nearly extremal MTZ black hole}

The form of the function $f(r)$ does not allow us to find an analytical solution for $r_+$. Since the case $r_+=Q/2$ corresponds to extremal black holes, it is convenient to parametrize a nearly extremal black hole by
\begin{equation}
r_+=\frac{Q}{2}(1+\epsilon^{\prime}),
\end{equation}
where $\epsilon^{\prime}$ parametrises the closeness of the black hole to extremality. For a nearly extremal black hole $\epsilon^{\prime}$ is considerably smaller than unity; whereas it is identically zero for an extremal black hole. We substitute this value in the equation $f(r_+)=0$. Using
\[
\ln(r_+^2)=\ln\left[ \left( \frac{Q}{2}\right)^2 \right] +2\epsilon^{\prime} -(\epsilon^{\prime})^2,
\]
One can get
\[
\left( \frac{Q}{2} \right)^2 +2(\epsilon^{\prime})^2 \left( \frac{Q}{2} \right)^2 -M-\left( \frac{Q}{2} \right)^2\ln\left( \frac{Q}{2} \right)^2=0,
\]
which implies
\begin{eqnarray}
\delta_{\rm{in}}&=&M-\left( \frac{Q}{2} \right)^2\left[ 1- \ln \left(\frac{Q}{2} \right)^2 \right] \nonumber
\\&=& 2(\epsilon^{\prime})^2 \left( \frac{Q}{2} \right)^2. \label{deltain}
\end{eqnarray}
We start with a nearly extremal black hole with $\delta_{\rm{in}}$, given by (\ref{deltain}). Again we demand $\delta_{\rm{fin}}<0$ so that the nearly extremal black hole is overcharged. We proceed the same way as the extremal case to derive the maximum value of $\delta E$ for a test particle with charge $\delta Q=Q\epsilon$ so that the black hole is overcharged at the end of the interaction. In order to find  $\delta E_{\rm{max}}$ for nearly extremal black holes, one can substitute Eq. (\ref{deltain}) in Eq. (\ref{deltafin1}) which leads to
\begin{eqnarray}
\delta E_{\rm{max}}&=&-(\epsilon^2+2\epsilon)\left( \frac{Q}{2}\right)^2\ln\left[\left( \frac{Q}{2}\right)^2\right]\nonumber\\&&-\Big[2\epsilon^2+2(\epsilon^{\prime})^2\Big] \left( \frac{Q}{2}\right)^2 .\label{deltafin3}
\end{eqnarray}
The behaviour of maximum energy is similar as shown in Fig. 1. For a numerical example, let us choose $(Q/2)^2=0.5$, and $\epsilon=\epsilon^{\prime}=0.01$. Using $\delta_{\rm{in}}=2(\epsilon^{\prime})^2(Q/2)^2$, we find that $M=0.84667$. (\ref{deltafin3}) implies that $E_{\rm{max}}=0.006766$. Let us choose $\delta E=0.001$ which satisfies the condition $E_{\rm{min}}<E<\delta E_{\rm{max}}$. ( Note that $E_{\rm{min}}=Q(\delta Q) \ln (r_+)$ is still negative.)  $\delta_{\rm{fin}}$ is given by
\begin{eqnarray}
 \delta_{\rm{fin}}&=&M+\delta E-\left( \frac{Q+\delta Q}{2}\right)^2 \nonumber\\&&+ \left( \frac{Q+\delta Q}{2}\right)^2\ln\left[ \left( \frac{Q+\delta Q}{2}\right)^2 \right]\nonumber\\&=&-0.00577.
\label{exnearext}
\end{eqnarray}
The negative sign for $\delta_{\rm{fin}}$ shows that nearly extremal black holes can also be overcharged.

\subsection{Backreaction effects}

We have shown that an extremal and a nearly MTZ black hole could be overcharged. We would like to check the hypothesis whether could a near extremal black hole be overcharged or not if one takes all the second order perturbations into account. Here we follow the work of Sorce and Wald  \cite{Sorce-Wald17}, where the authors argued that the violation of WCCC for nearly extremal black holes can be fixed by considering all non-linear order perturbations. In other words, in this section, we adapt their method to check the overcharging of MTZ black holes by considering all the second order perturbations.  Let's now recall Eq.~(\ref{Eq:extremal}),
$$
\delta\equiv M-\left( \frac{Q}{2} \right)^2\left[ 1- \ln\Big( \frac{Q}{2}\Big)^2 \right] \, .
$$
The cases with $\delta>0$ represent black hole solutions while the cases with $\delta <0$ correspond to objects without an event horizon. Let us consider the small deviations from the initial value of $\delta$, by expressing it as a one-parameter family of perturbation function with small parameter $\lambda$, i.e.
\begin{eqnarray}\label{Eq:one-parameter}
\delta(\lambda)\equiv M(\lambda)-\left( \frac{Q(\lambda)}{2} \right)^2\left[ 1- \ln\Big( \frac{Q(\lambda)}{2}\Big)^2 \right]\, ,
\end{eqnarray}
with $M(\lambda)$ and $Q(\lambda)$ given by 
\begin{eqnarray}
M(\lambda)&=& M+\lambda \delta E\, , \\
Q(\lambda)&=& Q+\lambda \delta Q\, . 
\end{eqnarray}
Here we choose $\delta E$ and $\delta Q$ in such a way that they are in agreement with the first order optimal perturbation.    

Note that for a nearly extremal black hole, $\delta(0)=2(\epsilon^{\prime})^2 (Q/2)^2$ is given by Eq. (\ref{deltain}). The terms linear in $\lambda$ correspond to the first order perturbations. We now expand $\delta(\lambda)$ up to second order in  $\lambda$ to study the effect of the second order perturbations, which also include the backreactions,
\begin{eqnarray}
\delta(\lambda) &=&  \frac{1}{2} Q^2 (\epsilon^{\prime})^2 +\lambda \left[ \delta E + Q\delta Q \ln \left(\frac{Q}{2} \right) \right] \nonumber \\
&&+ \frac{\lambda^2}{2} \left[ \delta^2 E +\delta Q^2 + \delta Q^2 \ln \left(\frac{Q}{2} \right)\right. \nonumber\\&&+ \left. Q \ln \left(\frac{Q}{2} \right) \delta^2 Q \right]
+O(\lambda^3). \label{lambdaexpand}
\end{eqnarray}
In the method of Sorce and Wald, the terms $\delta^2 E$ and $\delta^2 Q$ represent the backreaction effects. Now we need to substitute the expression for  $\delta E$ for the optimal perturbation. The optimal perturbation corresponds to the minimum value for $\delta E$ which allows the absorption of the test particle/field, as defined by Sorce and Wald. In the nearly extremal MTZ case the optimal perturbation satisfies (See Eq. (\ref{eminopt}) )
\begin{eqnarray}
\delta E&=& Q \delta Q\ln (r_+)=Q \delta Q\ln \left(\frac{Q(1+\epsilon^{\prime})}{2}  \right) \nonumber \\
 &=& Q \delta Q\ln \left(\frac{Q}{2}  \right) + Q\delta Q \epsilon^{\prime} +O\left((\epsilon^{\prime})^2\right) \label{deltaEcorrect}
\end{eqnarray}
This is the minimum value for the energy of the particle to be absorbed by the black hole.  Note that the small parameter $\epsilon^{\prime}$ determines the black hole's closeness to extremality, and the case $\epsilon^{\prime}=0$ corresponds to extremal black holes. With this substitution Eq. (\ref{lambdaexpand}) takes the form
\begin{eqnarray}
\delta(\lambda) &=&  \frac{1}{2} Q^2 (\epsilon^{\prime})^2 +\lambda \left[2 Q\delta Q \ln \left(\frac{Q}{2} \right)+Q\delta Q\epsilon^{\prime} \right] \nonumber \\
&&+\frac{\lambda^2}{2} \left[ \delta^2 E +\delta Q^2 + \delta Q^2 \ln \left(\frac{Q}{2} \right)\right. \nonumber \\ &&+ \left. Q \ln \left(\frac{Q}{2} \right) \delta^2 Q \right] 
+O(\lambda^3). \label{deltafinlambda}
\end{eqnarray}
{Since the second order terms $\delta^2 E$ and $\delta^2 Q$ are of the order $O(\epsilon^2)$}~\cite{Sorce-Wald17}, in this case, the expression in (\ref{deltafinlambda}) is not positive definite. For example for $Q^2=1$, $\delta Q=Q\epsilon$, one derives
\begin{equation}
\delta (\lambda)\sim \frac{(\epsilon^{\prime})^2}{2}- 0.69 \lambda \epsilon +\lambda \epsilon \epsilon^{\prime} +O(\lambda^2 \epsilon^2).
\label{deltafinlambdafin}
\end{equation}
Eq. (\ref{deltafinlambdafin}) implies that $\delta (\lambda)$ will be negative for $\lambda\sim \epsilon$. Therefore the second order perturbations cannot compensate for the overcharging of nearly extremal MTZ black holes. In this work we have ignored the gravitational radiation effects. However one would naively expect the contribution of the backreaction effects to be of the order $O(\epsilon^2)$. In that case, Eq. (\ref{deltafinlambda})  implies  that the contribution of the second order terms will be of the order $O(\epsilon^2)$, which will not contribute to $\delta(\lambda)$. Thus, we have shown that in dimensions $d<4$ MTZ black holes could be overcharged irrespective of whether the second order perturbations are included or not. This allows us to understand the nature of the MTZ black hole better in dimensions $d<4$ as its horizon is not stable as compared to the one in four dimensions. We do not attempt to apply the same method for extremal black holes, since the result will be negative definite, with $\delta (0)=0$.

\section{Conclusion}

In this paper, we have investigated the validity of the weak cosmic censorship conjecture for the charged MTZ black hole. We evaluated the cases of both the extremal and near-extremal black holes. In Wald type problems one derives a minimum and a maximum energy for the particles. If the energy of the particle is less than the minimum energy $\delta E_{\rm{min}}$, the particle is not absorbed by the black hole. On the other hand if the energy is larger than $\delta E_{\rm{max}}$ the black hole cannot be overcharged. If $\delta E_{\rm{min}}<\delta E_{\rm{max}}$, there exists a range of energies which allows us to overcharge black holes into naked singularities. For the extremal MTZ black hole, we have only considered the solutions with $M<1$ and $(Q/2)^2<1$ for which overcharging is possible. We have  shown that, the $(2+1)$ dimensional charged black holes dissociate from their four and higher dimensional analogues in two respects: The minimum energy at the horizon is negative, and extremal black holes can be overcharged. Nearly extremal $(2+1)$ dimensional black holes can also be overcharged similar to the (3+1) and higher dimensional ones in the case when the backreaction effects  are ignored.  

 The question then arises -- what happens for the charged MTZ black hole when backreaction effects are included? To address it, we have also incorporated the backreaction effects into our analysis adapting the second order corrections defined by Sorce and Wald~\cite{Sorce-Wald17}. However, we have shown that the inclusion of second order corrections cannot prevent the nearly extremal MTZ black holes from being overcharged. Hence it turns out that the overcharging of MTZ black holes in dimension $d<4$ appears to be rather generic. This is an interesting aspect of the charged MTZ black hole in dimension $d<4$ that refuses what is true for black holes in four dimension. 

 We have mentioned that the backreaction effects, which are represented by the $\delta^2 E$ terms in the Sorce-Wald method, will not contribute to $\delta(\lambda)$ if they are of the order $O(\epsilon^2)$, as one would expect in the case of test particles. Moreover, equations (\ref{deltafinlambda}) and (\ref{deltafinlambdafin}) imply that their contribution will be negligible even if $\delta^2 E \sim \epsilon$.  This suggests that it could be more appropriate to explicitly calculate the gravitational radiation and electromagnetic self-force effects and incorporate into the calculation of $\delta_{\rm{fin}}$, rather than adapting the Sorce-Wald method. In that case the numerical examples (\ref{exext}), and (\ref{exnearext}) suggest that the backreaction effects can make $\delta_{\rm{fin}}$ positive and preclude the violation of cosmic censorship if their contribution is of the order $O(\epsilon)$. It is not possible to observe this effect in the Sorce-Wald method. However, we should also note that it seems unlikely that the backreaction effects  make a high contribution changing the current scenario. Typically, the contribution of the backreaction effects is expected to be of the order of $\epsilon^2$.

\bibliographystyle{apsrev4-1}  
\bibliography{gravreferences}

\end{document}